\documentclass[%
 aip,
 amsmath,amssymb,
 reprint,%
]{revtex4-1}

\usepackage{graphicx}
\usepackage{dcolumn}
\usepackage{bm}

\usepackage[utf8]{inputenc}
\usepackage[T1]{fontenc}
\usepackage{mathptmx}
\usepackage{etoolbox}
\usepackage{color}
\makeatletter
\def\@email#1#2{%
 \endgroup
 \patchcmd{\titleblock@produce}
  {\frontmatter@RRAPformat}
  {\frontmatter@RRAPformat{\produce@RRAP{*#1\href{mailto:#2}{#2}}}\frontmatter@RRAPformat}
  {}{}
}%
\makeatother
\begin{document}

\preprint{AIP/123-QED}

\title[]{
Lifshitz-like Metastability and Optimal Dephasing in Dissipative Bosonic Lattices}
\author{Stefano Longhi}
 \email{stefano.longhi@polimi.it}
 \altaffiliation[Also at ]{IFISC (UIB-CSIC), Instituto de Fisica Interdisciplinar y Sistemas Complejos - Palma de Mallorca, Spain}
\affiliation{ Dipartimento di Fisica, Politecnico di Milano, Piazza L. da Vinci 32, I-20133 Milano, Italy}
%


\date{\today}

\begin{abstract}
In dissipative bosonic systems, dephasing is typically expected to accelerate relaxation and suppress coherent dynamics. However, we show that in networks of coherently coupled bosonic modes with non-uniform local dissipation, the presence of quasi-dark states leads to a nontrivial response to dephasing: while weak dephasing facilitates equilibration, moderate to strong dephasing induces a pronounced slowdown of relaxation, revealing the existence of an optimal dephasing rate that enhances equilibration. Using exact dynamical equations for second-order moments, we demonstrate that dephasing suppresses coherent transport and gives rise to long-lived collective modes that dominate the system's late-time behavior. This phenomenon bears striking similarities to Lifshitz-tail states, which are known in disordered systems to cause anomalously slow relaxation. Our results uncover a counterintuitive mechanism by which dephasing, rather than promoting equilibration, can dynamically decouple specific modes from dissipation, thereby protecting excitations. These findings highlight how non-Hermitian physics in open bosonic systems can give rise to unexpected dynamical regimes, paving the way for new strategies to control relaxation and decoherence in bosonic quantum systems, with broad implications for both experimental and theoretical quantum science.
\end{abstract}
\maketitle

\section{Introduction}
Bosonic lattices and networks---such as photonic lattices, optomechanical arrays, ultracold atoms in optical lattices and superconducting quantum circuits---have emerged as powerful platforms for simulating nonequilibrium quantum dynamics in engineered environments~\cite{R1,R2,R2b,R3,R4,R5}.
These systems are not only key platforms for quantum simulation but also play a central role in the rapidly developing field of non-Hermitian physics, where effective non-Hermitian Hamiltonians are used to model the interplay between coherent dynamics, loss and gain~\cite{R6,R7}. This framework has led to the discovery of a range of exotic phenomena, including exceptional points and parity-time symmetry breaking~\cite{R8,R8b,R8c,R8d,R8f,R9,R10,R10b}, the {non-Hermitian skin effect~\cite{R11,R12,R13,R14,R15,R14b,R14c,R14d,R14e,R14f}, non-Hermitian topological phases and phase transitions~\cite{R14,R15,R15b,R16,R17,R17b,R17c,R17d,R17e}, anomalous transport behavior \cite{R18,R19,R20,R21,R22}, and new mechanisms for controlling localization~\cite{R23}. However, to fully capture quantum effects such as decoherence, entanglement, syncrhonization and quantum-to-classical transition a complete open quantum systems description \cite{R24,R25} is required (see e.g. \cite{R25b,R25c,R26,R27,R27b,R28,R29,R30,R31,LonghiOL}).
 In this broader framework, relaxation dynamics can display intriguing features \cite{R27,R28,R32,R33,R34,R34b}, which are of major importance in understanding thermalization, metastability, and coherence preservation in open quantum systems. In models with spatially localized dissipation or when dissipation occurs within the same environment, certain collective excitations--- such as {quasi-dark states} or metastable modes \cite{R37b}---can become weakly coupled to the environment, giving rise to long-lived metastable dynamics~\cite{R32,R35,R36,R37,R37b,LonghiJMO}. These modes often preserve nonzero coherences that decay only at long timescales, leading to a pronounced separation of relaxation times \cite{R32}.  Dephasing, which suppresses phase coherence between system eigenstates, is generally expected to {accelerate} relaxation by destroying quantum interference responsible for dark or quasi dark-states, allowing the system to more quickly explore its available states. This behavior has been observed in various contexts, where dephasing assists thermalization and can enhance energy transport (see e.g. \cite{R38,R39,R40,R41,R42,R43,R44}). In particular, dephasing typically drives a breakdown of genuine quantum coherence and induces a transition from coherent quantum dynamics to effectively classical transport behavior, although certain features of non-Hermitian coherent dynamics may persist under specific conditions\cite{LonghiLST}.
 
In this research article we show that in networks of coherently coupled bosonic modes with non-uniform local dissipation, the presence of quasi-dark states leads to a nontrivial response to dephasing: while weak dephasing facilitates equilibration, moderate to strong dephasing induces a pronounced slowdown of relaxation, reminiscent of the behavior observed in the Zeno regime of strong dissipation \cite{R45}. We demonstrate that dephasing suppresses coherent transport and dynamically decouples certain collective modes from the dissipative channels. These long-lived modes, dominating the late-time dynamics and resulting into a significantly slower relaxation than in the purely dissipative case, exhibit features reminiscent of \emph{Lifshitz tail states}~\cite{R46,R47,R47b,R48,R49,R50,R50b,R51}, which are known to govern anomalously slow dynamics in disordered systems. This leads to an optimal dephasing rate that facilitates faster equilibration, balancing the suppression of quasi-dark states while avoiding the slow dynamics associated with isolated metastable Lifshitz tail states. Our results
uncover a counterintuitive mechanism by which dephasing, instead of accelerating equilibration, can under certain conditions protect excitations and inhibit relaxation. This mechanism offers new opportunities for engineering metastable dynamics in open quantum systems, with broad relevance for bosonic quantum technologies.

\section{Dissipative bosonic networks: theoretical model}
We consider a rather arbitrary linear network of  bosonic modes comprising $N$ sites or nodes, such as a bosonic lattice, with local dissipative terms describing single-particle loss and gain processes, with rates $\gamma_n$ and $g_n$, respectively  ($n$ is node number), and with uniform dephasing at a rate $\Gamma$ [Fig.1(a)]. Under Markovian dynamics, the evolution of the density operator $\rho(t)$ of the bosonic field is governed by the Lindblad master equation \cite{R24,R25,R51b,R52,R53}
$ (d \rho/dt)= \mathcal{L} \rho(t)$ with Liouvillian
\begin{eqnarray}
\mathcal{L} \rho & = & -i [ H, \rho]+\sum_n \left(  \gamma_n \Lambda [a_n]+g_n \Lambda [a^{\dag}_n]+ \Gamma \Lambda [a_n^{\dag} a_n] \right) 	\nonumber 
\\ & \equiv  & -i [H, \rho]+ \mathcal{D} \rho 
\end{eqnarray}
where  $H= \sum_{n,m} J_{n,m} a^{\dag}_n a_m$
describes the coherent Hamiltonian dynamics, with $J_{n,m}=J_{m,n}^*$, $a_n$, $a^{\dag}_n$ are the destruction and creation bosonic operators at network node $n$ satisfying the usual commutation relations 
\[ [a_n, a^{\dag}_m] = \delta_{n,m} \;, \;\;  [a_n, a_m] = [a_n^{\dag}, a^{\dag}_m]=0 \], 
and 
\begin{equation}
\Lambda[o]= o \rho o^{\dag}-(1/2) (o^{\dag} o \rho+ \rho o^{\dag} o) \nonumber
\end{equation} 
are the dissipative superoperators, with the jump operators $o={a}_n$, $o={a}^{\dag}_n$ and $o={a}_n^{\dag}a_n$ describing local particle loss, gain and dephasing at rates $\gamma_n$, $g_n$ and $\Gamma$, respectively.  The Liouvillian superoperator can be equivalently written as
\begin{eqnarray}
\mathcal{L} \rho & = & -i \left( {H}_{eff} \rho-\rho {H}_{eff}^{\dag}  \right)+ \sum_n \gamma_n a_n \rho a_n^{\dag} \nonumber \\
& + & \sum_n g_n a^{\dag}_ n \rho a_n+ \Gamma \sum_ n a^{\dag}_n a_n \rho a^{\dag}_n a_n
\end{eqnarray}
where
\begin{equation}
{H}_{eff}={H}-\frac{i}{2}  \sum_n \left(  \gamma_n a^{\dag}_na_n+g_n a_n a^{\dag}_n+ \Gamma a^{\dag}_n a_n a^{\dag}_n a_n  \right) \nonumber
\end{equation}
is an effective non-Hermitian Hamiltonian,  and the last three terms on the right hand side of Eq.(2), $\sum_k {L}_k \rho {L}_{k}^{\dag}$, describes quantum jumps.
To avoid instabilities, we assume $g_n< \gamma_n$. We also assume that there is a single stationary (equilibrium) state $\rho_0$, such that $\mathcal{L} \rho_0=0$. When $\Gamma=0$, i.e. in the absence of dephasing, the Liouvilllian is quadratic  in the bosonic operators, the  equilibrium state is Gaussian and its existence and unicity can be proved under rather general conditions \cite{R54,R55,R56,R57}. For example, taking $g_n / \gamma_n= \exp(-\beta)$ independent of site index $n$, it can be readily shown that the stationary state is given by $\rho_0=Z^{-1} \exp(- \beta \sum_{n} a^{\dag}_n a_n)$, where $Z$ is the normalization factor and the parameter $\beta$ provides the strength of thermal excitation. Remarkably, this Gaussian state is also the equilibrium state when considering dephasing ($\Gamma \neq 0$), even though the Liouvillian is not anymore quadratic.   
We consider the interesting situation where, besides the stationary state, the system admits of quasi dark states \cite{R37b}, i.e. slowly-decaying (metastable) states, when dephasing is switched off ($\Gamma=0$).  A dark state is a pure state $\rho_d=| \psi_d \rangle \langle \psi_d|$  where $| \psi_d \rangle$ is an eigenstate of $H$ such that $\mathcal{D} \rho_d=0$. $\rho_d$ is clearly an eigenstate of the Liouvillian $\mathcal{L}$ as well.  For a quasi-dark state, instead, $\mathcal{D} \rho_d \simeq 0$ and $\rho_d$ becomes a resonance \cite{R37b}. Dark and quasi-dark states arise rather generally in networks under certain topology or symmetry constraints, such that some eigenstates of $H$ have not excitations in dissipative nodes of the network (see for instance \cite{LonghiOL,R37b}). Indicating by $\xi_n$ ($n=1,2,..,N$) an eigenstate of the $N \times N$ matrix $J_{n,m}$ with $\xi_n=0$ at some nodes $n \in \mathcal{S}$, let us assume $\gamma_n=g_n=0$ for $n \not\in \mathcal{S}$; then any Fock state $| \psi_d \rangle$, obtained from the vacuum state by application of the creation operator $b^{\dag} = \sum_n \xi_n a^{\dag}_n$, is an eigenstate of $H$ that does not dissipate and is thus a dark state. A quasi-dark state is obtained by introducing a small loss rate at an additional site $n=n_0 \not\in \mathcal{S}$, which yields metastability in the relaxation process toward the equilibrium state. For example, in a linear lattice with open boundary conditions and nearest-neighbor hopping rate $J$ [Fig.1(b)], comprising an odd number of sites $N$ with the bosonic modes possessing the same frequency, the eigenstates of the tridiagonal matrix $J_{n,m}$ are given by the standing waves $\xi_n \propto \sin [\pi l n/(N+1) ]$ ($l,n=1,2,...,N$). For $l=(N+1)/2$, $\xi_n$ vanishes at every even site of the lattice, $\mathcal{S}= \{ 2,4,...,N-1\}$, and thus dark states arise whenever $\gamma_n=g_n=0$ at odd sites.

\begin{figure}
\includegraphics[width=8.5 cm]{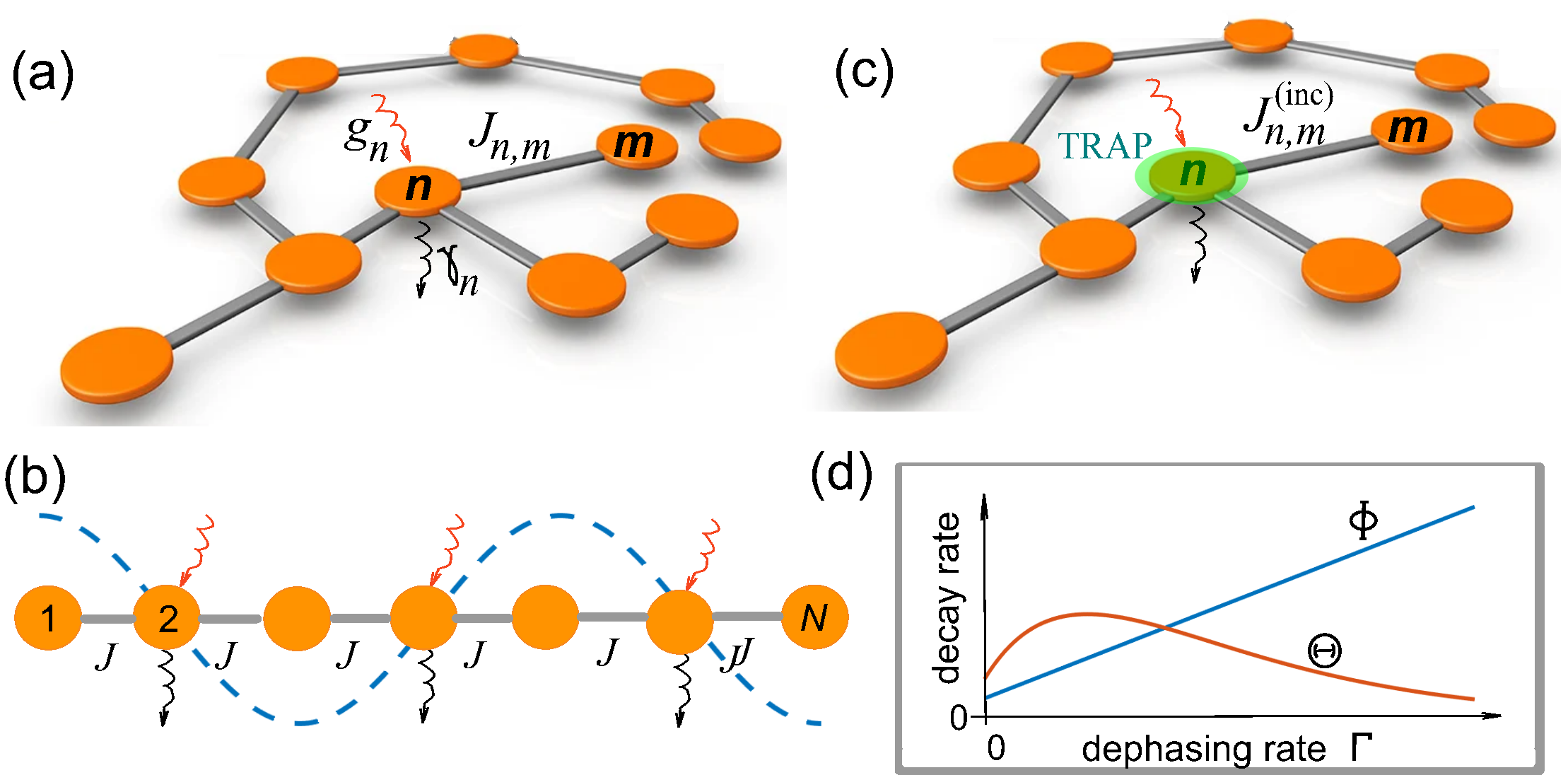}
\caption{(a) Schematic of a quantum bosonic network. The coherent hopping rate between nodes $n$ and $m$ is $J_{n,m}$ ($n \neq m$), $J_{n,n}$ are the resonance frequencies. 
Some nodes of the network, belonging to the subset $n \in \mathcal{S}$, can dissipate into separate baths. Local dissipative terms describe single-particle loss and gain processes, with rates $\gamma_n$ and $g_n$, respectively. (b) A one-dimensional tight-binding lattice of bosonic modes with open boundary conditions, nearest-neighbor coherent hopping amplitude $J$ and equal resonance frequencies. The single-particle eigenstates of the lattice are standing waves (dashed curve). If the dissipative sites $n \in \mathcal{S}$ are nodes (zeros) of the standing wave, then the system shows dark states. A quasi-dark state is obtained by adding a small loss term to an additional site where the standing wave has a non-vanishing excitation. (c) When large dephasing at a rate $\Gamma$ is uniformly added to each node of the network, the system behaves classically and the dynamics of the mean values of excitations (populations) $P_n(t)$  is governed by the random walk of a classical particle on the network with traps, the incoherent hopping rates between nodes $n$ and $m$ being given by $J_{n,m}^{(inc)}=2|J_{n,m}|^2/ \Gamma$. The classical particle can survive for long times in the dissipation-free islands of the network (metastable states) before being annihilated by a trap. (d) Sketch of the expected behavior of the relaxation rates $\Phi$ and $\Theta$ of first- and second-moments versus dephasing rate $\Gamma$. In the strong dephasing region (classical regime) a characteristic slow-down of relaxation for $\Theta$ is expected owing to the formation of long-lived metastable states (Lifshitz tail states). 
This leads to an optimal dephasing rate that facilitates faster equilibration, balancing the suppression of quasi-dark states while avoiding the slow dynamics associated with isolated metastable Lifshitz tail states.}
\end{figure}

\section{Relaxation dynamics}
The central question is to elucidate the role of dephasing in the relaxation process toward equilibrium, which is notably hindered by the presence of quasi-dark states. In our model, the Lindbladian and effective non-Hermitian approaches exhibit significantly different dynamical behaviors, primarily because the jump operators representing dephasing effects are quadratic in the bosonic operators \cite{R21}. To accurately capture the impact of dephasing on the emergence of long-lived Lifshitz tail states, it is therefore essential to explicitly incorporate quantum jumps into the dynamical description.\\
 When the relaxation dynamics does not display strong non-normal features \cite{R30,R31,R57b}, the relaxation time is basically established by the inverse of the Liouvillian spectral gap. For $\Gamma=0$, the Liouviallian $\mathcal{L}$ is quadratic  in the bosonic creation/destruction operators and the model is fully integrable via the third-quantization method \cite{R54,R55}. In particular, the entire dynamics is fully described by the equations of motion of first-order and second-order moments \cite{R29,R51b,R57}, and the spectral gap of the Liouvillian $\mathcal{L}$ is given by the decay rate of the slowest-decaying mode in the relaxation dynamics of the covariance matrix \cite{R55}.  Therefore, to characterize the relaxation dynamics toward the equilibrium state we can limit considering the relaxation dynamics of the mean values $A_n(t)={\rm Tr} ( \rho(t) a_n)$ and of the two-point correlation matrix $C_{n,m}(t)={\rm Tr} ( \rho(t) a_n^{\dag} a_m)$. 
In the presence of dephasing, the Liouvillian is quartic in the bosonic operators, however an extension of the third-quantization method is still possible \cite{R56,R57} and a closed-form equation for $A_n(t)$ and $C_{n,m}(t)$ can be obtained as well. For our network model defined by the Liouvillian given in Eq.(1), the evolution equations for $A_n(t)$ and $C_{n,m}(t)$ read explicitly
\begin{eqnarray}
\frac{dA_n}{dt}  & = &  -i \sum_m J_{n,m} A_m(t)-\frac{1}{2} (\gamma_n+\Gamma-g_n) A_n  \nonumber \\
& \equiv & \sum_{m} \mathcal{A}_{n,m} A_m(t) 
\end{eqnarray}
\begin{eqnarray}
\frac{dC_{n,m}}{dt} & = & -i \sum_l \left( J_{m,l} C_{n,l}-J_{l,n}C_{l,m}   \right)- \Gamma (1-\delta_{n,m}) C_{n,m} \nonumber \\
& - & \frac{1}{2} \left( \gamma_n+\gamma_m-g_n-g_m \right) C_{n,m}  +g_n \delta_{n,m} \nonumber \\
& \equiv & \sum_{l_1,l_2} \mathcal{B}_{n,m; l_1,l_2} C_{l_1,l_2}+G_{n,m} 
\end{eqnarray}
where we have set $G_{n,m} \equiv g_n \delta_{n,m}$ and where $\mathcal{A}$ and $\mathcal{B}$ are $N \times N$ and $N^2 \times N^2$ matrices. The equilibrium values are readily obtained by letting $dA_n/dt=0$ and $dC_{n,m}/dt=0$ in the above equations. Clearly, the equilibrium value $(A_n)_{eq}$ for the mean values $A_n$ is $(A_n)_{eq}=0$. On the other hand, the equilibrium value $(C_{n,m})_{eq}$ for the second moments is non-vanishing owing to the forcing term $G_{n,m}$ due to incoherent pumping. When $g_n/ \gamma_n= \exp(-\beta)$,  $(C_{n,m})_{eq}$ is diagonal, independent of $\Gamma$ and $n$, and given by $(C_{n,m})_{eq}=g_n \delta_{n,m} /(\gamma_n-g_n)=\delta_{n,m} / (\exp(\beta)-1)$ [see as an example Fig.2(a)], showing the characteristic divergence of the correlations as the instability point $\beta \rightarrow 0^+$ is approached. On the other hand, when the ratio $g_n / \gamma_n$ depends on $n$, the stationary correlation matrix $(C_{n,m})_{eq}$ is not anymore diagonal and its elements depend on $\Gamma$ [an example is given in Fig.3(a)].

The eigenvalues $\lambda_{\alpha}$ and $\mu_{\alpha,\beta}$ of the two matrices $\mathcal{A}$ and $\mathcal{B}$, with $\alpha,\beta=1,2,...,N$, determine the relaxation rates toward the equilibrium. In the stable regime $g_n < \gamma_n$, all eigenvalues have negative real part.  We indicate by $\Phi (\Gamma)={\rm min}_{\alpha} | {\rm Re} (\lambda_{\alpha})|$ and $\Theta(\Gamma)={\rm min}_{\alpha,\beta} | {\rm Re}(\mu_{\alpha,\beta})|$ the eigenvalues of the two matrices with the smallest value (in absolute value) of the real part, corresponding to the slowest decaying rates in the system. When the dephasing is switched off, i.e. for $\Gamma=0$, one has $\mu_{\alpha,\beta}= \lambda_{\alpha}+\lambda_{\beta}$,  $\Theta=2 \Phi$  and $\Phi$ is very small, owing to the existence of long-lived quasi-dark states. As $\Gamma$ is increased above zero, clearly from the form of $\mathcal{A}$ it is expected that $\Phi (\Gamma)$  increases monotonously and secularly with $\Gamma$, i.e. the mean values $A_n(t)$ undergo a faster damping toward the equilibrium value $(A_n)_{eq}=0$ as a result of the additional dephasing [Fig.1(d)]. Likewise, $\Theta ( \Gamma)$ is expected to increase as well since dephasing destroys phase coherence and thus the quasi-dark (metastable) states in the system. However, for large dephasing rates the decay rate $\Theta(\Gamma)$ is expected to {\em decrease} and vanish as $\Gamma \rightarrow \infty$, as sketched in Fig.1(d). The underlying physical mechanism is that strong dephasing induces diffusive (incoherent) transport within the network, which can be effectively described by a classical Markov master equation. In the limit of large dephasing strength $\Gamma$, the hopping rate vanishes, and the system enters the so-called Zeno regime \cite{R45}. In this regime, new metastable states can emerge within dissipation-free regions, reminiscent of Lifshitz tail states, which are known to govern anomalously slow dynamics in disordered systems \cite{R46,R47,R47b,R61,R62}. This leads to an optimal dephasing rate [Fig.1(d)] that facilitates faster equilibration, balancing the suppression of quasi-dark states while avoiding the slow dynamics associated with isolated metastable Lifshitz tail states.

 To clarify this point, let us consider the dynamical equations (4) in the large dephasing regime, where $\Gamma$ is much larger than the coherent hopping rates $|J_{n,m}|$ and dissipative rates $\gamma_n,g_n$, i.e. $( |J_{n,m}|, \gamma_n, g_n ) / \Gamma \sim \epsilon \ll 1$. In this regime, the correlations $C_{n,m}(t)$ for $n \neq m$ rapidly decay to small values, at least of order $ \sim \epsilon$, and they can be eliminated adiabatically from the dynamics using standard perturbative methods (see for example \cite{R58,R59,R60}). The relevant equation for the diagonal elements $P_n(t)=C_{n,n}(t)={\rm Tr} ( \rho(t) a^{\dag}_n a_n)$, i.e. of mean excitation numbers (populations), is given by the classical master equation
\begin{equation}
\frac{dP_n}{dt}= \sum_{m \neq n} J^{(inc)}_{n,m}(P_m-P_n)-(\gamma_n-g_n)P_n+ g_n
\end{equation}
where $J^{(inc)}_{n,m}=J^{(inc)}_{m,n}=2|J_{n,m}|^2 / \Gamma$ is the effective incoherent hopping rate between nodes $n$ and $m$ of the network. After letting $P_n(t)=P^{(eq)}_n+p_n(t)$, where
$P_n^{(eq)}$ 
is the equilibrium state, Eq.(5) reads
\begin{equation}
\frac{dp_n}{dt}= \sum_{m \neq n} J^{(inc)}_{n,m}(p_m-p_n)-(\gamma_n-g_n)p_n.
\end{equation}
The above equation formally describes a continuous-time classical random walk on the network with traps placed at the dissipative nodes, $n \in \mathcal{S}$ [Fig.1(c)]: whenever the particle sits at the dissipative node $n$, it is annihilated at the rate $(\gamma_n-g_n)$ per unit time. Trapping problems of this kind have been studied extensively (see e.g. \cite{R47b,R61,R62}), and the asymptotic longtime falloff of the survival probability, establishing the time scale of relaxation toward equilibrium, is known to be related to Lifshitz tail states \cite{R47b,R50,R50b,R61}. These are long-lived metastable states where the classical particle resides for long times in wide dissipation-free regions of the network before being annihilated by the traps. 

\section{Illustrative examples and discussion}
\subsection{\textcolor{black}{Linear chain with determinist loss/gain}}
As a \textcolor{black} {first} illustrative example, let us consider a linear chain (a lattice) of $N$ sites with equal resonance frequencies, nearest-neighbor hopping amplitude $J$ and open boundary conditions [Fig.1(b)],  where dissipation takes uniform values $\gamma_n=\gamma$, $g_n=g$ in a fraction $p$ of sites $n \in \mathcal{S}$, while $\gamma_n=g_n=0$  in the other fraction $(1-p)$ of sites $n \not\in \mathcal{S}$. In this case, in the large $N$ limit the decay rate $\Theta(\Gamma)$ of the most long-lived Lifshitz  tail state is given by \cite{R50b}
\begin{equation}
\Theta(\Gamma) \simeq \frac{2 \pi^2 J^2}{\Gamma (l_m+1)^2}
\end{equation}
where  $l_m$ is longest island of consecutive dissipation-free sites in the lattice. As one can see from Eq.(7), the relaxation rate of $P_n(t)$ vanishes as $ 1 / \Gamma$ in the larger dephasing (classical) regime. \textcolor{black}{To obtain a quasi-dark state, it is sufficient to apply the gain and loss terms in some even sites of the lattice, with an additional weak loss/gain term in one site of the lattice with odd number. For example, let us assume a lattice comprising} $N=21$ sites, and let us assume $\mathcal{S}=\{ 2,6,8,20 \}$ as the set of dissipative sites, with $\gamma_n /J=1.2 $ and $g_n/J=0.2$ at such sites. \textcolor{black}{This lattice sustains an exact dark state. To obtain a quasi-dark state, an}  additional small dissipation is introduced at the odd site $n_0=1$, with $\gamma_{n_0}/J=0.06$ and $g_{n_0}/J=0.01$, which transforms dark states into long-lived quasi-dark states at $\Gamma=0$. For such parameter values, one has $l_m=11$ for the longest island of consecutive dissipation-free sites in the lattice. Since $g_n/\gamma_n=\exp(-\beta)=1/6$ is independent of $n$, the stationary values of correlations $(C_{n,m})_{eq}$ are diagonal and independent of $\Gamma$ and $n$, as depicted in Fig.2(a). Figures 2(b) and (c) show the numerically-computed behavior of the relaxation rates $\Phi$ and $\Theta$ versus $\Gamma$, as obtained by diagonalization of the matrices $\mathcal{A}$ and $\mathcal{B}$ entering in Eqs.(3) and (4). According to the theoretical predictions, while the relaxation rate $\Phi$ for the mean values $A_n(t)$ monotonously increases as $\Gamma$ is increased [Fig.2(b)], the relaxation rate $\Theta$ of correlations undergoes an initial increase with $\Gamma$, with a largest decay rate at around $\Gamma / J \simeq 0.25 $ [inset in Fig.2(c)], corresponding to the optimal dephasing value for fastest equilibration. Above this value, $\Theta$ decreases with $\Gamma$ and vanishes as $\Gamma / J \rightarrow \infty$. The dashed curve in Fig.2(c) depicts the theoretically-predicted value of the decay rate $\Theta$ in the large dephasing regime of classical random walk [Eq.(7)], which fits very well with the exact numerical results. 

\begin{figure}
\includegraphics[width=8.8 cm]{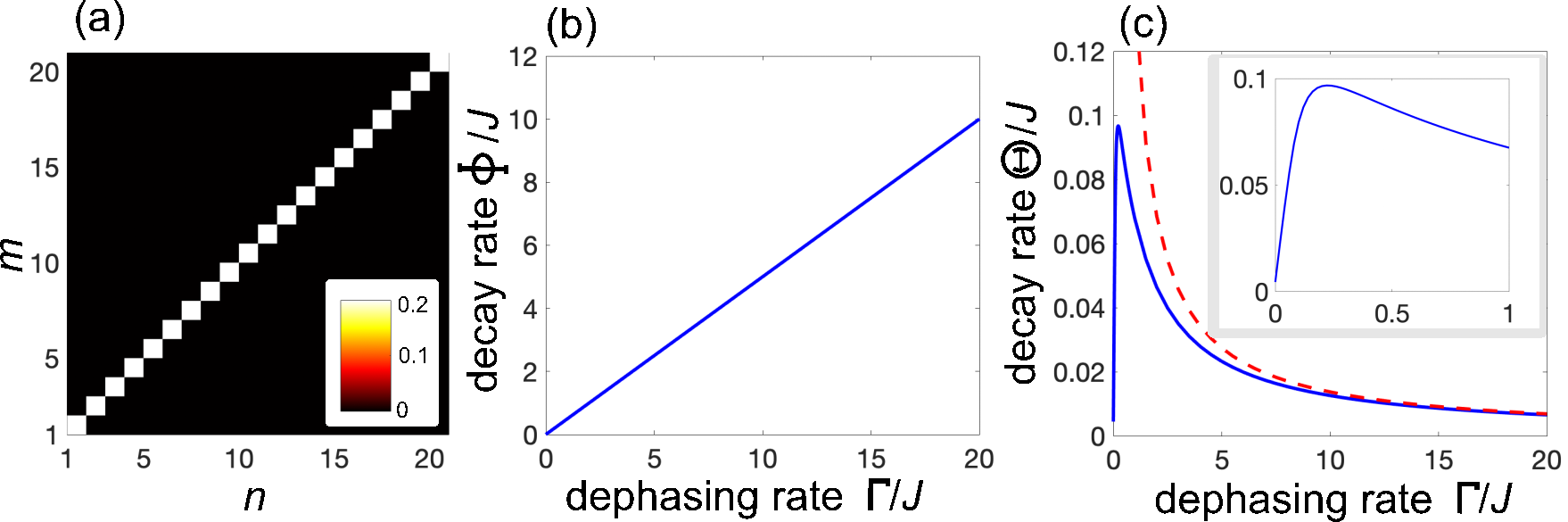}
\caption{(a) Equilibrium state of the single-particle correlation matrix (modulus of $(C_{n,m})_{eq}$ on a pseudocolor map) in a linear chain of bosonic oscillators with the same resonance frequencies, nearest neighbor hopping rate $J$ and open boundary conditions [Fig.1(b)]. The lattice comprises $N=21$ sites. The set of dissipative sites is  $\mathcal{S}=\{ 2,6,8,20 \}$ , with $\gamma_n/J=1.2$, $g_n/\gamma_n=1/6$ for $n \in \mathcal{S}$ and $\gamma_n=g_n=0$ otherwise. An additional small dissipation is placed at site $n=n_0=1$ to create quasi dark states in the dephasing-free regime, namely $\gamma_{n_0}/J=0.06$ and $g_{n_0}/\gamma_{n_0}=1/6$. Since the ratio $g_n/ \gamma_n$ is independent of $n$, the equilibrium state is diagonal, does not depend on site index $n$ neither on dephasing rate $\Gamma$. (b,c) Numerically-computed behavior of the relaxation rates of mean values $A_n(t)$ and correlations $C_{n,m}(t)$, $\Phi$ and $\Theta$, versus the dephasing rate $\Gamma$. The dashed red curve in (c) shows the behavior of the decay rate of correlations as predicted in the classical random walk limit $\Gamma / J \gg 1$ [Eq.(7)]. The inset in (c) depicts an enlargement of the decay rate $\Theta$ versus $\Gamma$ in the weak dephasing regime, clearly demonstrating the existence of an optimal value $\Gamma / J \simeq 0.25$ at which equilibration is faster.}
\end{figure}

Since the formation of long-lived Lifshitz tail states requires solely the existence of islands of dissipation-free nodes, the above scenario is observed even when the ratio $ g_n/\gamma_n$ depends on index $n$, i.e. when the equilibrium correlation matrix $(C_{n,m})_{eq}$ is not diagonal and its elements depend on $\Gamma$. An illustrative example is shown in Fig.3. In this example we consider the same linear chain of Fig.2 with the same parameter values, except than the condition $ g_n/\gamma_n= \exp(-\beta)=1/6$ is broken by letting  $g_8/ \gamma_8=1/3$  at the dissipative node $n=8$. Figure 3(a) depicts three indicative behaviors of the stationary correlation matrix   $(C_{n,m})_{eq}$ for increasing values of dephasing rate $\Gamma$, showing that for vanishing or small dephasing the correlation matrix is not diagonal. The behavior of decay rates $\Phi$ and $\Theta$ versus dephasing rate $\Gamma$ is plotted in Figs.3(b) and (c), clearly indicating the slow down of relaxation for the correlations at large dephasing rates due to the emergence of long-lived Lifshitz tail states. 

\begin{figure}
\includegraphics[width=8.8 cm]{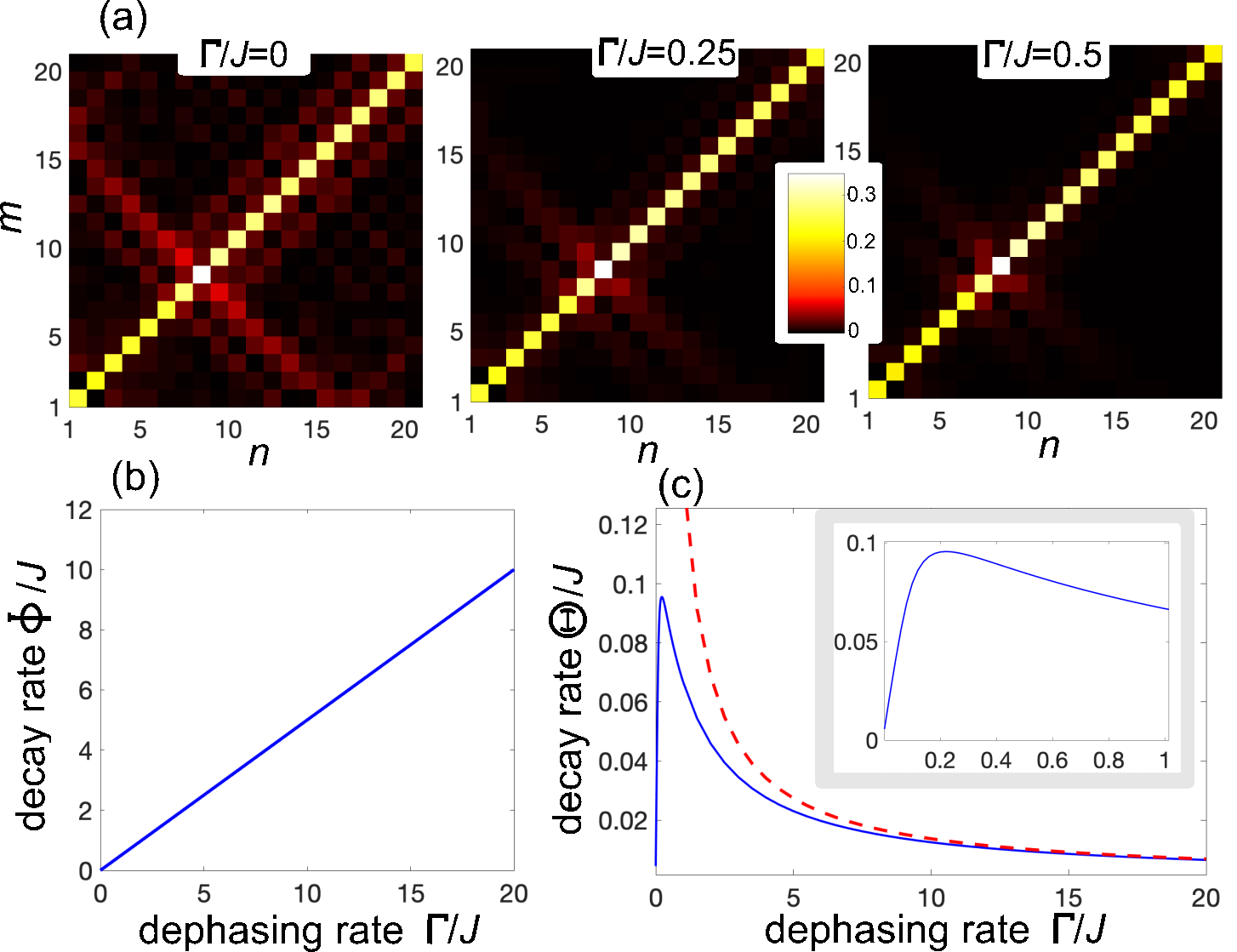}
\caption{Same as Fig.2, but for $g_8/ \gamma_8=1/3$  at the dissipative node $n=8$. In this case the condition $ g_n/\gamma_n= \exp(-\beta)=1/6$ is broken at site $n=8$ and the equilibrium state of the single-particle correlation matrix depends on the dephasing rate $\Gamma$, as indicated by the three panels in (a).}
\end{figure}

\textcolor{black}{
\subsection{Linear chain with stochastic gain/loss}
The previous example focused on deterministic distributions of dissipation that support quasi-dark states, however similar phenomena can also emerge when dissipation is applied to a subset $ \mathcal{S}$ of nodes selected stochastically within the network}. A notable example is the Bernoulli-Anderson model, where the local dissipation rates $\gamma_n$ are randomly assigned, taking the value $\gamma_n=\gamma$ with probability $p$, and $\gamma_n = 0$ with probability ($1-p)$. The gain rates are taken as $g_n= \gamma_n \exp(-\beta)$.
 In this case, due to the stochastic placement of dissipative nodes within the network, the formation of long-lived quasi dark states arising from destructive interference in the $\Gamma = 0$ limit is generally inhibited. Nevertheless, in the weak dephasing regime and for small concentrations of dissipative nodes, i.e. for  $p \ll 1$, equilibration can still be accelerated by dephasing, resulting in an optimal dephasing rate that provides the fastest relaxation to equilibrium. As dephasing increases above such an optimal value, the system crosses over into a classical hopping regime, where a pronounced slowdown of relaxation emerges, driven by the appearance of long-lived Lifshitz-tail-like states.
This behavior is illustrated in Figs.~4(a,b), which show the numerically computed decay rates $\Phi$ and $\Theta$ as functions of $\Gamma$ for a Bernoulli-Anderson distribution of dissipation in the same linear lattice considered in Fig.~2. The results are obtained by averaging over 100 stochastic realizations of dissipative node configurations, with parameter values $\gamma = 1.2J$, $p = 0.2$, and $g_n = \gamma_n/6$. In the classical limit of large dephasing, $\Gamma / J \gg 1$, the decay rate $\Theta$ is well approximated by Eq.~(7), where the characteristic length $l_m$ is replaced by its statistical average $\overline{l}_m$, computed using extreme value theory for stochastic Bernoulli processes~\cite{R61}
\begin{equation}
\overline{l}_m=- \frac{\ln (\sigma p N)}{\ln (1-p)},
\end{equation}
where $\sigma \sim 1$ is a form factor. As the concentration $p$ of dissipative nodes in the network increases toward $p = 1$, corresponding to the case where all nodes are dissipative, dephasing in the weak regime becomes progressively less effective in accelerating equilibration. In the limit $p \rightarrow 1$, the decay rate $\Theta$ turns into a monotonically decreasing function of $\Gamma$. This behavior is illustrated in Figs.~4(c) and 4(d).

\begin{figure}
\includegraphics[width=8.5 cm]{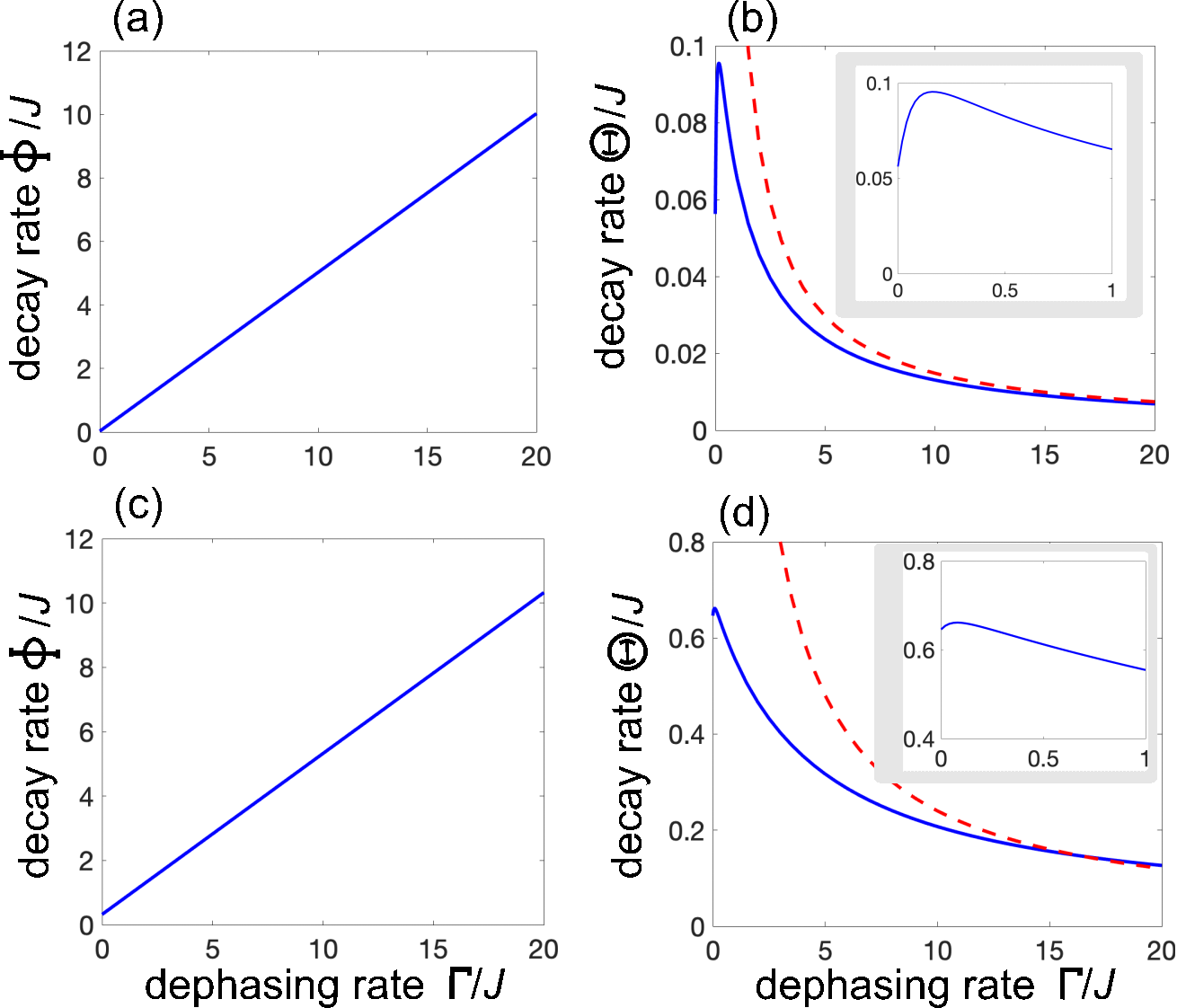}
\caption{(a,b) Numerically-computed behavior of the decay rates $\Phi$ [panel (a)] and $\Theta$ [panel (b)] versus the dephasing rate $\Gamma$ in the linear chain of Fig.1(b), where the set $\mathcal{S}$ of dissipative nodes are taken from a Bernoulli-Anderson distribution with concentration $p=0.2$. The curves are obtained after a statistical average over 100 different realizations of dissipation node distributions. Parameter values are given in the text. The dashed red curve in (b) displays the behavior of the decay rate predicted in the classical hopping limit of strong dephasing using Eqs.(7) and (8) with a form factor $\sigma=2.5$. The inset in (b) depicts an enlargement of the decay rate of correlations in the weak dephasing regime. Note that, as compared to the case of Fig.2(c), the decay rate $\Theta$ is larger at $\Gamma=0$ owing to the low probability to have dark or quasi-dark states for a stochastic distribution of dissipative nodes. However, an optimal value of the rate  $\Gamma$ that maximizes the decay rate is still clearly observed. (c,d) Same as (a,b), but for a concentration $p=0.8$ of dissipative nodes.}
\end{figure}

\textcolor{black}{
\subsection{Experimental Realizations and Platforms}
The theoretical framework developed in this work can be explored experimentally using several state-of-the-art platforms that support engineered bosonic modes with controlled dissipation and dephasing.  Promising candidates include nanophotonic systems such as coupled microcavities or photonic crystal cavities, optical waveguide lattices, synthetic photonic lattices, ultracold atomic gases, and superconducting quantum circuits.\\
A particularly accessible realization of the linear network in Fig.~1(b) is through arrays of evanescently coupled optical waveguides ~\cite{Longhi2009},  coupled microcavities or photonic crystal cavities ~\cite{Lodahl2015,Garanovich2012}, or coupled microring resonators ~\cite{Hafezi2013,R17b}.  In such systems, individual control of optical gain and loss can be achieved via spatially selective pumping, absorption, or scattering, enabling site-resolved engineering of dissipative processes \cite{R17b}. Dephasing can be introduced by modulating the refractive index (and thus the propagation constants) with noise in time or space, or by incoherent probing using broadband light sources~\cite{Caruso2009, Crespi2013}. These photonic systems provide excellent control and measurement capabilities, particularly suited for monitoring non-Hermitian transport dynamics.\\
Synthetic photonic lattices -- based on modulated ring resonators, fiber loops, or time-multiplexed quantum walk architectures ~\cite{Lustig2019, Yuan2018,Szameit2024}-- offer another versatile platform. In these systems, temporal degrees of freedom emulate spatial lattices, and gain/loss elements and phase modulators can be inserted to realize tailored Lindbladian evolution, including both local dissipation and dephasing \cite{LonghiLST,R28,dynamicdisorder}. These setups enable dynamic control over system parameters and have been used to simulate non-Hermitian physics and topological effects.\\
Ultracold atomic systems in optical lattices also present a viable setting. Localized loss and gain processes can be engineered using tightly focused electron beams or optical potentials, while dephasing can be introduced via controlled noise in lattice depths or by coupling to auxiliary thermal reservoirs~\cite{Barontini2013, Tomita2017}. Such systems inherently feature bosonic statistics and tunable interactions, providing a natural platform for realizing dissipative many-body quantum dynamics.		
Finally, superconducting quantum circuits offer chip-based realizations of coupled bosonic modes (e.g., microwave resonators or transmon qubits in the dispersive regime) with programmable loss, gain, and dephasing channels. Using quantum reservoir engineering techniques, these platforms have demonstrated precise control over open-system dynamics and have been used to explore non-Hermitian Hamiltonians and dissipative phase transitions~\cite{R42,Fitzpatrick2017, Ma2019}.
\subsection{Fermionic networks}
While the analysis presented in this work focuses on non-interacting bosons, similar considerations apply to networks of spinless, non-interacting fermions undergoing coherent hopping dynamics and subject to local particle gain/loss and dephasing processes \cite{fermions1}. In this case, the Lindblad master equation remains of the same general form as Eq.(1), but the creation and annihilation operators obey fermionic anti-commutation relations:
\begin{equation}
\left\{ a_n, a_m^{\dag} \right\} = \delta_{n,m}, \quad \left\{ a_n, a_m \right\} = \left\{ a_n^{\dag}, a_m^{\dag} \right\} = 0.
\end{equation}  
Owing to these different statistics, the evolution equation for the two-point correlation matrix, defined as $C_{n,m}(t) = \mathrm{Tr}[\rho(t)\, a_n^{\dag} a_m]$, takes a form distinct from its bosonic counterpart \cite{R56,R57,fermions1}. Specifically, one obtains:
\begin{eqnarray}
\frac{dC_{n,m}}{dt} & = & -i \sum_l \left( J_{m,l} C_{n,l} - J_{l,n} C_{l,m} \right) - \Gamma (1 - \delta_{n,m}) C_{n,m} \nonumber \\
& - & \frac{1}{2} \left( \gamma_n + \gamma_m + g_n + g_m \right) C_{n,m} + g_n \delta_{n,m} \nonumber \\
& \equiv & \sum_{l_1,l_2} \mathcal{B}^{(\mathrm{ferm})}_{n,m; l_1,l_2} C_{l_1,l_2} + G_{n,m},
\end{eqnarray}
where $\mathcal{B}^{(\mathrm{ferm})}$ denotes the generator of the dissipative dynamics and $G_{n,m}$ is the gain source term.
A key difference with respect to the bosonic case lies in the structure of the gain terms: for fermions, they appear with a damping sign in the relaxation matrix $\mathcal{B}^{(\mathrm{ferm})}$. This reflects the Pauli exclusion principle, which forbids more than one particle per site and thus prevents the occurrence of bosonic-like instabilities (i.e., unbounded population growth due to gain). As a consequence, the fermionic system is dynamically stable regardless of the gain/loss balance.
Aside from this statistical distinction -- which becomes negligible in the limit of weak or vanishing gain -- the overall relaxation dynamics of the covariance matrix, governed by the spectral properties of $\mathcal{B}^{(\mathrm{ferm})}$, exhibits behavior qualitatively similar to the bosonic case. In particular, weak dephasing suppresses quasi-dark modes and accelerates relaxation, while strong dephasing suppresses coherent hopping and leads to the emergence of long-lived, collective decay modes -- the Lifshitz tail states-- that dominate the system's late-time dynamics.
}

\section{Conclusions}
To conclude, this study uncovers a non-monotonic role of dephasing in the relaxation dynamics of bosonic networks with non-uniform dissipation, driven by the formation of metastable Lifshitz tail states. For weak dephasing rates, decoherence accelerates equilibration by disrupting the coherence in the system and destroying long-lived quasi-dark modes. However, at strong dephasing rates, a pronounced slowdown of relaxation is observed due to the emergence of long-lived metastable modes, reminiscent of Lifshitz tail states, which become decoupled from dissipation. This leads to an optimal dephasing rate that facilitates faster equilibration, balancing the suppression of quasi-dark states while avoiding the slow dynamics associated with isolated metastable Lifshitz tail states. \textcolor{black}{Similar results should occur for the relaxation dynamics of spinless non-interacting fermions in dissipative networks}. The findings demonstrate that dephasing, rather than always accelerating relaxation, can in some regimes inhibit decay and protect excitations, offering a new route for controlling relaxation and decoherence in open quantum systems. The ability to tune dephasing could thus play a crucial role in engineering metastable states and controlling relaxation dynamics in a range of bosonic quantum systems, with significant implications for the development of robust quantum technologies and advancing the understanding of decoherence in modern quantum science.



\section*{Data Availability Statement}
Data available on request from the author.




\end{document}